# The Transactional Nature of Quantum Information

Subhash Kak

Department of Computer Science
Oklahoma State University
Stillwater, OK 74078

**ABSTRACT**
Information, in its communications sense, is a transactional property. If the received signals communicate choices made by the sender of the signals, then information has been transmitter by the sender to the receiver. Given this reality, the potential information in an unknown pure quantum state should be non-zero. We examine transactional quantum information, which unlike von Neumann entropy, depends on the mutuality of the relationship between the sender and the receiver, associating information with an unknown pure state. The information that can be obtained from a pure state in repeated experiments is potentially infinite.

**INTRODUCTION**

The term "information" is used with a variety of meanings in different situations. In the mathematical theory of communications [1], information is a measure of the surprise associated with the received signal. It is implied that the receiver has knowledge of the statistics of the messages produced, or to be produced, by the sender. A more likely signal carries less information as it comes with less surprise.

Let the sender and the receiver share a set of messages from a specific alphabet and the statistics of the communications will allow us to determine the probability of each letter of the alphabet. The information measure of the message $x$ associated with probability $p_x$ is $-\log p_x$. Classical informational entropy is given by

$$H(X) = -\sum_{x} p_x \log p_x \qquad (1)$$

where $p_x$ is the probability of the message $x$.

The amount of information associated with an object could be taken to mean the amount necessary to completely describe it. Since this information will vary depending on the interactions the object has with other objects and fields, and thus be variable, it may be measured for the situation where the object is isolated.

In the signal context information in a signal is sometimes seen from the lens of complexity [2],[3],[4]. On frequency considerations, the two patterns:

$$\begin{array}{l} 1\ 1\ 1\ 1\ 1\ 1\ 1\ 1\ 1\ 1\ 1\ 1 \\ 1\ 0\ 1\ 1\ 0\ 1\ 0\ 0\ 1\ 0\ 0\ 0\ 1 \end{array} \qquad (2)$$

assumed to be generated by tosses of a fair coin must be taken to be equally random. But from the perspective of complexity, the first pattern is less random than the second since it can be described more compactly.

One may speak of information associated with an object also from the perspective of uncertainty, either in the limitations related to its employment in a computational situation, or in defining a deeper symmetry with classical objects [5],[6]. Another view would be to see how much information can be carried by the object. The physical nature of information was emphasized in different ways by Wheeler and Landauer [7],[8],[9].

Let us consider now what is the information associated with a quantum object, say, a photon. Represented as a qubit $(a\,|\,0\rangle + b\,|\,1\rangle)$, the photon will collapse to $|\,0\rangle$ or $|\,1\rangle$ but since this collapse is random, it would not communicate any useful information to the receiver. Maximum information will be communicated to the receiver if the sender prepares the photon in one of the two orthogonal states, say, $|0\rangle$ or $|1\rangle$, which the receiver will be able to determine upon observation. It is assumed that the sender and the receiver have agreed upon the communication protocol and the measurement bases. If the agreement on the nature of the communication had not been made in advance, the receiver cannot, in general, obtain useful information from the stream of photons (excepting in their presence or absence) since he does not know the basis states of the sender, or know if these states are fixed or variable.

If he knew the basis states of the sender, he would obtain the maximum of one bit of information from each qubit.

Quantum information is traditionally measured by the von Neumann entropy, $S_n(\rho) = -tr(\rho \log \rho)$, where $\rho$ is the density operator associated with the state [10]. This entropy may be written also as

$$S_n(\rho) = -\sum_x \lambda_x \log \lambda_x \qquad (3)$$

where $\lambda_x$ are the eigenvalues of $\rho$. It follows that a pure state has zero von Neumann entropy (we will take "log" in the expression to base 2 implying the units are *bits*). The von Neumann measure is independent of the receiver. This is unsatisfactory from a communications perspective because if the sender is sending an unknown pure state to the receiver, it should, in principle, communicate information.

Suppose that the sender and the receiver have agreed that the sender will choose, say, one of 16 different polarization states of the photon. The sender picks one of these and sends several photons of this specific polarization. Since all these photons are the same pure state, the information associated with them, according to the von Neumann measure, is zero. But, the information in this choice, from the communications point of view, is $\log_2 16 = 4$ bits.

Although the classical entropy and the von Neumann entropy measures as given by equations (1) and (3) look similar mathematically, there is a difference between the two expressions, as the first one relates to probabilities and the second to the eigenvalues associated with a matrix.

## INFORMATION IN MIXED AND PURE STATES

A given photon may be in a pure or mixed quantum state, and these two cases are very different from the point of view of measurement. A mixed state is a statistical mixture of component pure states, and its entropy computed by the von Neumann measure is similar to the entropy for classical states. The maximum information provided by a single mixed state photon is one bit.

**Example 1.** Consider the mixed state $|\Psi\rangle = \frac{3}{4}|0\rangle\langle 0| + \frac{1}{4}|1\rangle\langle 1|$. Its von Neumann entropy equals 0.81 bits. This mixed state can be viewed to be generated from a variety of ensembles of states. For example, it can be viewed as the ensemble $|\Psi\rangle = \frac{1}{2}|a\rangle\langle a| + \frac{1}{2}|b\rangle\langle b|$, where $|a\rangle = \frac{3}{4}|0\rangle + \frac{1}{4}|1\rangle$ and $|b\rangle = \frac{3}{4}|0\rangle - \frac{1}{4}|1\rangle$. In each of these ensembles the entropy would be the same.

**Example 2.** Consider an entangled pair of quantum objects represented by the pure state $|\Psi\rangle = \frac{1}{\sqrt{2}}(|00\rangle + |11\rangle)$ that corresponds to the density operator

$$\rho = \begin{bmatrix} .5 & 0 & 0 & .5 \\ 0 & 0 & 0 & 0 \\ 0 & 0 & 0 & 0 \\ .5 & 0 & 0 & .5 \end{bmatrix} \qquad (4)$$

Its eigenvalues are 0, 0, 0, and 1, and, therefore, its von Neumann entropy is zero. But consider the two objects separately; their density operators are $\rho = \begin{bmatrix} .5 & 0 \\ 0 & .5 \end{bmatrix}$ each, which means that their entropy is 1 bit each.

We are now confronted with the situation that the two components of an entangled quantum system have non-zero entropy although the system taken as a whole has entropy of zero! Intuitively, this is not reasonable because one expects the information of a system to be function of the sum of its parts.

The zero entropy of the entangled state is unreasonable also from the perspective that it could have been in one of many different forms and therefore having information potential. For example, with the assumption that the probability amplitudes be real and equal, the state function could have been either $\frac{1}{\sqrt{2}}(|00\rangle + |11\rangle)$ or $\frac{1}{\sqrt{2}}(|00\rangle - |11\rangle)$, communicating information equal to one bit.

Information in classical theory measures the reduction in uncertainty regarding the message based on the received communication. It is predicated on the mutuality underlying the sender and the receiver. Therefore the receipt of an unknown pure state should communicate information to the recipient of the state.

In our examination of the question of information in an unknown pure state $|\Psi\rangle = a|0\rangle + b|1\rangle$, we take it that multiple copies of the state are available. The received copies can be used by the recipient estimate the values of $a$ and $b$. If the objective is the estimation of the density matrix $\rho$, we can do so by finding average values of the observables $\rho$, $X\rho$, $Y\rho$, $Z\rho$, where X, Y, Z are the Pauli operators, by means of the expansion:

$$\rho = \frac{1}{2}(tr(\rho)I + tr(X\rho)X + tr(Y\rho)Y + tr(Z\rho)Z) \tag{5}$$

The operators $tr(X\rho)$ etc are average values and therefore the probabilistic convergence of $\rho$ to its correct value would depend on the number of observations made.

Considering a more constrained setting, assume that the information that user A, the preparer of the states, is trying to communicate to the user B, is the ratio $a/b$, expanded as a numerical sequence, $m$. For further simplicity it is assumed that A and B have agreed that $a$ and $b$ are real, then $a = \frac{m}{\sqrt{1+m^2}}$ and $b = \pm\frac{1}{\sqrt{1+m^2}}$. One may use either a pure state $|\phi\rangle = a|0\rangle + b|1\rangle$ or a mixed state consisting of an equal mixture of the states $|\phi_1\rangle = a|0\rangle + b|1\rangle$ and $|\phi_2\rangle = a|0\rangle - b|1\rangle$.

Alternatively, one may communicate the ratio $a^2/b^2=m$. In this case, $a = \frac{\sqrt{m}}{\sqrt{1+m}}$ and $b = \pm\frac{1}{\sqrt{1+m}}$. Since $b^2 = \frac{1}{1+m}$, one needs to merely determine the sequence corresponding to the reciprocal of $1+m$ for the probability of the component state $|1\rangle$.

In yet another arrangement, the sender and the receiver may agree in advance to code the contents of the message in the probability amplitudes in some other manner. For example, they may agree on the following table relating binary sequence and the probability amplitude $a$:

| 000 | 0   |
|-----|-----|
| 001 | 1/8 |
| 010 | 2/8 |
| 011 | 3/8 |
| 100 | 4/8 |
| 101 | 5/8 |
| 110 | 6/8 |
| 111 | 7/8 |

Thus the binary sequence 101 will map to the qubit $\frac{\sqrt{5}}{\sqrt{8}}|0\rangle - \frac{\sqrt{3}}{\sqrt{8}}|1\rangle$.

The preparer of the quantum state may choose out of infinity of possibilities, and depending on the mutual relationship between the preparer and the receiver the pure state's information will vary from one receiver to another. In case the pure state chosen by the sender is aligned to the measurement basis of the receiver, the information transmitted will be zero.

In general, the information generated by the source equals the probability of choosing the specific state out of the possibilities available (this is the states *a priori* probability). If the set of choices is infinite, then the information generated by the source is unbounded. On the other hand, due to the probabilistic nature of the reception process, not all the information at the source is obtained at the receiver by the measurement. *The more the number of copies of the photon the receiver is sent, the more information will be extracted.* A single photon will of course communicate at best only one bit of information.

## PARALLEL BETWEEM CLASSICAL AND QUANTUM PROTOCOLS

We have spoken of sending information using a single pure quantum state using many of its copies. This may be compared, in the classical case, to the representation of the contents of an entire book by a single signal, *V*, obtained by converting the binary file of the book into a number by putting a "dot" before it to make it less than 1. If a large number of copies of this signal are sent, it is possible that the receiver could, in principle, determine its value even in the presence of noise, $\varepsilon$, taken to be symmetrically distributed about zero. This would be accomplished by summing a large number of received signals *W*, so that $E(W) = E(V) + E(\varepsilon) = V + E(\varepsilon)$ may be calculated accurately. As the number of copies available to be summed increases, the value $E(\varepsilon)$ goes to zero and

$$E(W) \to V$$

Although such a method is not practical for an entire file, more than one binary symbol of a message are coded into a single amplitude in non-binary systems.

If one wished to use a pure state to transmit information in several binary bits in a practical implementation, the transmission should contain several copies of the pure state to enable the receiver to extract the relevant information from it. If the pure state is used to transmit a single bit, then we have the case of the information being sent by choice between two alternatives.

## TRANSACTIONAL INFORMATION IN A QUANTUM STATE

In 2007, I proposed a measure of entropy that covers both pure and mixed states [11]. In this measure, which I call *quantum informational entropy*, $S_{inf}(\rho)$, the entropy is related only the diagonal terms of the density matrix:

$$S_{\inf}(\rho) = -\sum_i \rho_{ii} \log \rho_{ii} \qquad (6)$$

Let the density matrix be $2^n \times 2^n$, that is it is associated with *n* qubits. Then,

$$0 \leq S_{inf}(\rho) \leq n \qquad (7)$$

The proof that the least value of $S_{inf}(\rho)$ is zero is obvious when the transmitted state is pure and aligned to the computational basis of the receiver. The maximum value of $S_{inf}(\rho)$ is $n$ when the diagonal terms are equal.

Many properties of $S_{inf}(\rho)$ are similar to those of $S_n(\rho)$. For example, for the joint state $\rho^{AB}$ of quantum systems A and B,

$$S_{\inf}(A,B) \leq S_{\inf}(A) + S_{\inf}(B) \tag{8}$$

An example of this is the entangled state (Example 2), which has informational entropy of 1 bit and the informational entropy of each of the components is also 1 bit.

The entropy function satisfies the inequality:

$$S_{\inf}(\sum_i p_i \rho_i) \geq \sum_i p_i S_{\inf}(\rho_i) \tag{9}$$

The intuition behind this is that the knowledge of the ensemble that has led to the generation of the quantum state reduces the uncertainty and, therefore, the entropy of the right hand side is less. Due to the same reason, we can write that

$$S_{inf}(\rho) \geq S_n(\rho) \tag{10}$$

If the receiver has knowledge that the ensemble consists of a specific probabilistic combination of pure and mixed components, then the partial entropy, $Sp(\rho) = \sum_i p_i S_i(\rho_i)$, is lower compared to when he has no such knowledge.

**Example 3.** Consider $\rho = \begin{bmatrix} 0.71 & 0.15 \\ 0.15 & 0.29 \end{bmatrix}$.

The informational entropy $S_{inf}(\rho)$ is

$$S_{inf}(\rho) = -0.71 \log_2 .71 - 0.29 \log_2 .29 = 0.868 \text{ bits}.$$

The eigenvalues of $\rho$ are 0.242 and 0.758 and, therefore, the von Neumann entropy $S_n(\rho)$ for this case is equal to

$$S_n(\rho) = -0.242 \log_2 .242 - 0.758 \log_2 .758$$
$$= .242 \times 2.047 + .758 \times .400 = 0.798 \text{ bits}.$$

The informational entropy exceeds the von Neumann entropy by 0.07 bits.

*Partial Information Case 1.* Assume that partial information is available for Example 3, and we know that the ensemble consists of a pure and a mixed component in the following manner:

$$\rho = 0.3 \times \begin{bmatrix} .5 & .5 \\ .5 & .5 \end{bmatrix} + 0.7 \times \begin{bmatrix} .8 & 0 \\ 0 & .2 \end{bmatrix}$$

Then the partial entropy associated with it will be the sum of the pure and mixed entropies and this turns out to be equal to $0.3 + 0.7 \times 0.722 = 0.805$ bits.

*Partial Information Case 2.* On the other hand, if we are told that the ensemble consists of pure and mixed component in the following manner:

$$\rho = 0.316 \times \begin{bmatrix} \frac{2}{3} & \frac{\sqrt{2}}{3} \\ \frac{\sqrt{2}}{3} & \frac{1}{3} \end{bmatrix} + 0.684 \times \begin{bmatrix} 0.731 & 0 \\ 0 & 0.269 \end{bmatrix}$$

The partial entropy will now be equal to:

0.316×0.918+0.684×0.84=0.290+0.575=0.865 bits

As in the previous case, in this partial measure, entropy has two components: one informational (related to the pure components of the quantum state), and the other that is thermodynamic (which is receiver independent).

For a two-component elementary mixed state, the most information in each measurement is one bit, and each further measurement of identically prepared states will also at most be one bit. For an unknown pure state, the information in it represents the choice the source has made out of the infinity of choices related to the values of the probability amplitudes with respect to the basis components of the receiver's measurement apparatus. Each measurement of a two-component pure state will provide at most one bit of information, and if the source has made available an unlimited number of identically prepared states the receiver can obtain additional information from each measurement until the probability amplitudes have been correctly estimated. Once that has occurred, unlike the case of a mixed state, no further information will be obtained from testing additional copies of this pure state.

## QUANTUM CRYPTOGRAPHY

The BB84 [10] and the three-stage quantum cryptography protocols [12] may be seen as emerging naturally from the different perspectives of computing signal operations on two sets of bases (mixed state scenario) and a single set (for each user).

## DISCUSSION

The approach of this paper is consistent with the positivist view that one cannot speak of information associated with a system excepting in relation to an experimental arrangement together with the protocol for measurement. The experimental arrangement is thus integral to the amount of information that can be obtained and it means that information in a unknown pure state is non-zero.

The receiver can make his estimate by adjusting the basis vectors so that he gets *closer* to the unknown pure state. The information that can be obtained from such a state in repeated experiments is potentially infinite in the most general case.

A distinction may be made in the situation for discrete (and finite) and continuous geometries. Since the measure of information in a pure state is a consequence of its "distance" from the observer's computational basis, information in a continuous geometry would for almost each observer (excepting for the one who starts out exactly aligned, the probability of which is zero) remain infinite. Conversely, for a finite discrete geometry, a smaller "distance" will translate into lesser information, making some observers more privileged than others.

**REFERENCES**

[1] Shannon, C.E., "A mathematical theory of communication." Bell System Technical Journal 27: 379-423, 623-656 (1948).

[2] Solomonoff, R., "A formal theory of inductive inference." Information and Control 7: 1-22, 224-254 (1964).

[3] Kak, S., "Classification of random binary sequences using Walsh-Fourier analysis." IEEE Trans. On Electromagnetic Compatibility EMC-13, 74-77 (1970).

[4] Li, M. and Vitanyi, P., An Introduction to Kolmogorov Complexity and its Applications. Springer Verlag (2008).

[5] Kak, S., "On quantum numbers and uncertainty", Nuovo Cimento 33B, pp. 530-534 (1976).

[6] Kak, S., "On information associated with an object", Proceedings Indian National Science Academy 50, pp. 386-396 (1984).

[7] Wheeler, J., "It from bit", Proceedings 3$^{rd}$ International Symposium on Foundations of Quantum Mechanics, Tokyo (1989).

[8] Landauer, R., "The physical nature of information", Phys. Lett. A 217, 188- 193 (1996).

[9] Kak, S., "Information, physics and computation", Foundations of Physics 26, 127-137 (1996).

[10] Nielsen, M.A. and Chuang, I.L., Quantum Computation and Quantum Information. Cambridge University Press (2000).

[11] Kak, S., "Quantum information and entropy", International Journal of Theoretical Physics 46, 860-876 (2007).

[12] Kak, S., "A three-stage quantum cryptography protocol." Foundations of Physics Letters 19, 293-296 (2006).